    \documentclass[%
 aip,
 amsmath,amssymb,
 reprint,%
]{revtex4-1}

\usepackage{graphicx}
\usepackage{dcolumn}
\usepackage{bm}
\usepackage{hyperref}
\usepackage[utf8]{inputenc}
\usepackage[T1]{fontenc}
\usepackage{mathptmx}
\usepackage{etoolbox}

\makeatletter
\def\@email#1#2{%
 \endgroup
 \patchcmd{\titleblock@produce}
  {\frontmatter@RRAPformat}
  {\frontmatter@RRAPformat{\produce@RRAP{*#1\href{mailto:#2}{#2}}}\frontmatter@RRAPformat}
  {}{}
}%
\makeatother
\begin{document}

\preprint{AIP/123-QED}

\title{Beyond One-Thousandth Energy Resolution with an
AlMn TES Detector}
\author{Liangpeng Xie}
\affiliation{School of Physics and Materials, Nanchang University, Nanchang,330031, China.}
 
\author{Yifei Zhang}
\email{zhangyf@ihep.ac.cn.}
\affiliation{State Key Laboratory of Particle Astrophysics, Institute of High Energy Physics, Chinese Academy of Sciences, Beijing 100049, China}

\author{Zhengwei Li}
\affiliation{State Key Laboratory of Particle Astrophysics, Institute of High Energy Physics, Chinese Academy of Sciences, Beijing 100049, China}

\author{Zhouhui Liu}
\affiliation{State Key Laboratory of Particle Astrophysics, Institute of High Energy Physics, Chinese Academy of Sciences, Beijing 100049, China}

\author{Shibo Shu}
\affiliation{State Key Laboratory of Particle Astrophysics, Institute of High Energy Physics, Chinese Academy of Sciences, Beijing 100049, China}

\author{Junjie Zhou}
\affiliation{School of Physics and Materials, Nanchang University, Nanchang,330031, China.}

\author{Xufang Li}
\affiliation{State Key Laboratory of Particle Astrophysics, Institute of High Energy Physics, Chinese Academy of Sciences, Beijing 100049, China}

\author{Haoyu Li}
\affiliation{State Key Laboratory of Particle Astrophysics, Institute of High Energy Physics, Chinese Academy of Sciences, Beijing 100049, China}

\author{He Gao}
\affiliation{State Key Laboratory of Particle Astrophysics, Institute of High Energy Physics, Chinese Academy of Sciences, Beijing 100049, China}

\author{Yudong Gu}
\affiliation{State Key Laboratory of Particle Astrophysics, Institute of High Energy Physics, Chinese Academy of Sciences, Beijing 100049, China}

\author{Xuefeng Lu}
\affiliation{State Key Laboratory of Particle Astrophysics, Institute of High Energy Physics, Chinese Academy of Sciences, Beijing 100049, China}

\author{Yong Zhao}
\email{zhaoyong@ncu.edu.cn.}
\affiliation{School of Physics and Materials, Nanchang University, Nanchang,330031, China.}

\author{Congzhan Liu}
\email{liucz@ihep.ac.cn.}
\affiliation{State Key Laboratory of Particle Astrophysics, Institute of High Energy Physics, Chinese Academy of Sciences, Beijing 100049, China}

\date{\today}

\begin{abstract}
The superconducting Transition-Edge Sensor (TES) is a critical technology for next-generation X-ray spectrometers, known for its exceptional energy resolution. In the last decade, TESs based on AlMn alloy films have been extensively used in several cosmic microwave background (CMB) experiments. The advantages of simple fabrication process and easily tunable critical temperature make them an alternative to bilayer TESs. However, they have rarely been applied to X-ray detection until now. We developed an annular AlMn TES for X-ray detection and tested it in a dilution refrigerator with a Superconducting Quantum Interference Device (SQUID) amplifier, achieving an Full Width at Half Maximum (FWHM) of 12.1 $\pm$ 0.3 eV at 17.48 keV. To the best of our knowledge, this is the first demonstration of an AlMn TES achieving an energy resolution below 0.1\%, highlighting its potential for high-resolution X-ray detection.

\end{abstract}

\maketitle

The superconducting Transition-Edge Sensor, as a high-sensitivity non-dispersive detector \cite{Irwin-2005}, has been widely employed in applications such as CMB detection \cite{Stevens-2020,Ding-2017}, nuclear physics \cite{Yoho-2020,Koehler-2021}, and material science experiments at synchrotron beamline facilities \cite{Uhlig-2015}. It has also been proposed for next generation X-ray astronomy space missions, including HUBS \cite{Cui-2020} and ATHENA \cite{Pajot-2018}. State-of-the-art TES microcalorimeter has achieved an ultra-high energy resolution of 1.6 eV at 5.9 keV (Mn K$\alpha$) \cite{Smith-2012}, which is nearly two orders of magnitude better than semiconductor detectors such as SDD. Since the 1990s, bilayer films such as Mo/Au, Ti/Au and Mo/Cu have been widely used in TES devices, where the critical temperature ($T_c$) can be tuned via the proximity effect. In recent years, TES devices based on AlMn alloy films have emerged as a promising alternative, since the critical temperature can be easily tuned via an annealing method reported in Reference \cite{Li-2016}. Compared with bilayer TESs, AlMn TESs offer a simpler fabrication progress, reduced sensitivity to magnetic fields, and more straightforward tunability of the critical temperature \cite{Vavagiakis-2018,Li-2016,Yu-2021,Liu-2025}.
These advantages have already resulted in the AlMn film being widely applied in newly building telescopes for CMB detection, such as SPT-3 G \cite{Anderson-2020}, POLARBEAR-2\cite{Westbrook-2018}and AliCPT-1\cite{Salatino-2021}. However, until now it has rarely been used for X-ray detection. Over the past few years, we have worked to extend its application to astronomical X-ray mission, such as the proposed Wide-band X-ray Polarization Telescope (WXPT)\cite{Yin-2022,Zhang-2025}. In previous studies, we developed several X-ray TES detectors based on AlMn alloy film and achieved an energy resolution of 11.0 eV at 5.9 keV \cite{Zhang-2025}, demonstrating the potential of AlMn TES for X-ray detection. 
In this work, we report the first achievement of a relative energy resolution below 0.1\% (12.1 eV at 17.48 keV) in an AlMn TES, establishing it as viable alternative to conventional bilayer TES devices for high-resolution X-ray spectroscopy.

\begin{figure}
\includegraphics[width=0.45\textwidth]{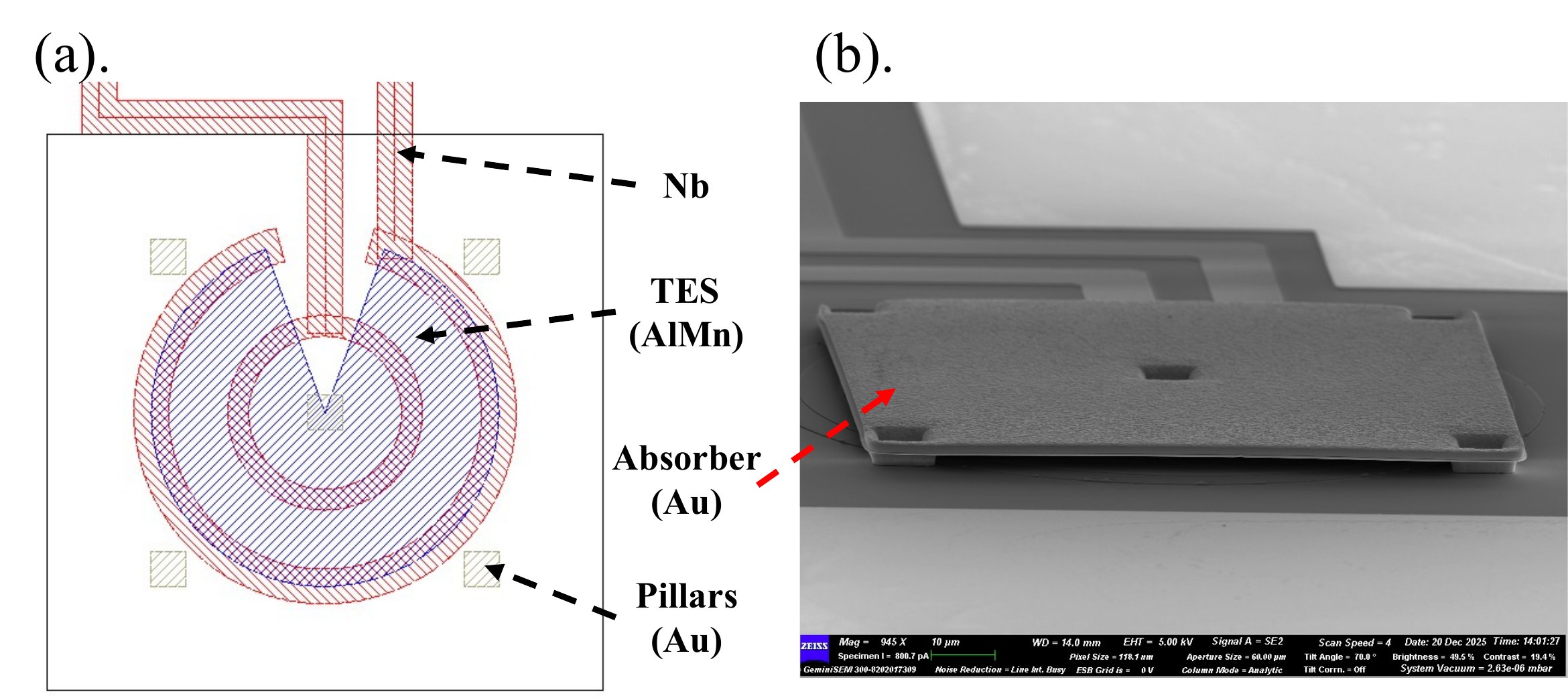} 
\caption{\label{fig11_layout_SEM} 
(a) Schematic of the TES device comprising two Nb electrodes and an AlMn film on a Si$_3$N$_4$/SiO$_2$ membrane (square area). The gold absorber is not shown for clarity. (b) SEM image of the TES detector fabricated.}
\end{figure}

Fig.~\ref{fig11_layout_SEM} shows the TES layout and SEM image of the newly developed TES detector. The AlMn TES had an annular shape with inner radius of 28 $\mu$m and outer radius of 45 $\mu$m. The thickness of the AlMn TES film was 300 nm. The inner Nb electrode, with a width of 10 $\mu$m and a thickness of 120 nm, was directly connected through a 40° notch in the annular TES, rather than being routed through the SiO$_2$ isolation layer above the AlMn film as in our previous work \cite{Zhang-2025}. The fabrication sequence of them was reversed here. The Nb electrodes were first deposited by DC magnetron sputtering and patterned using dry etching, followed by deposition of the AlMn film via DC magnetron sputtering using a new AlMn target with 2000 ppm Mn. Prior to TES patterning, the AlMn film was annealed on a hotplate at 230°C for 10 minutes to tune its critical temperature near 100 mK, basing on the calibrated Tc - annealing temperature curve. The gold absorber had an area of 100 $\mu$m by 100 $\mu$m, with a thickness of 1.7 $\mu$m. It was suspended directly above the annular AlMn TES by five gold pillars. Each pillar is 10 $\mu$m wide and 2.2 $\mu$m high. The central pillar serves as a thermal link between the absorber and the TES. More detailed design and fabrication procedures are described in Reference \cite{Zhang-2025}.

\begin{figure}
\includegraphics[width=0.45\textwidth]{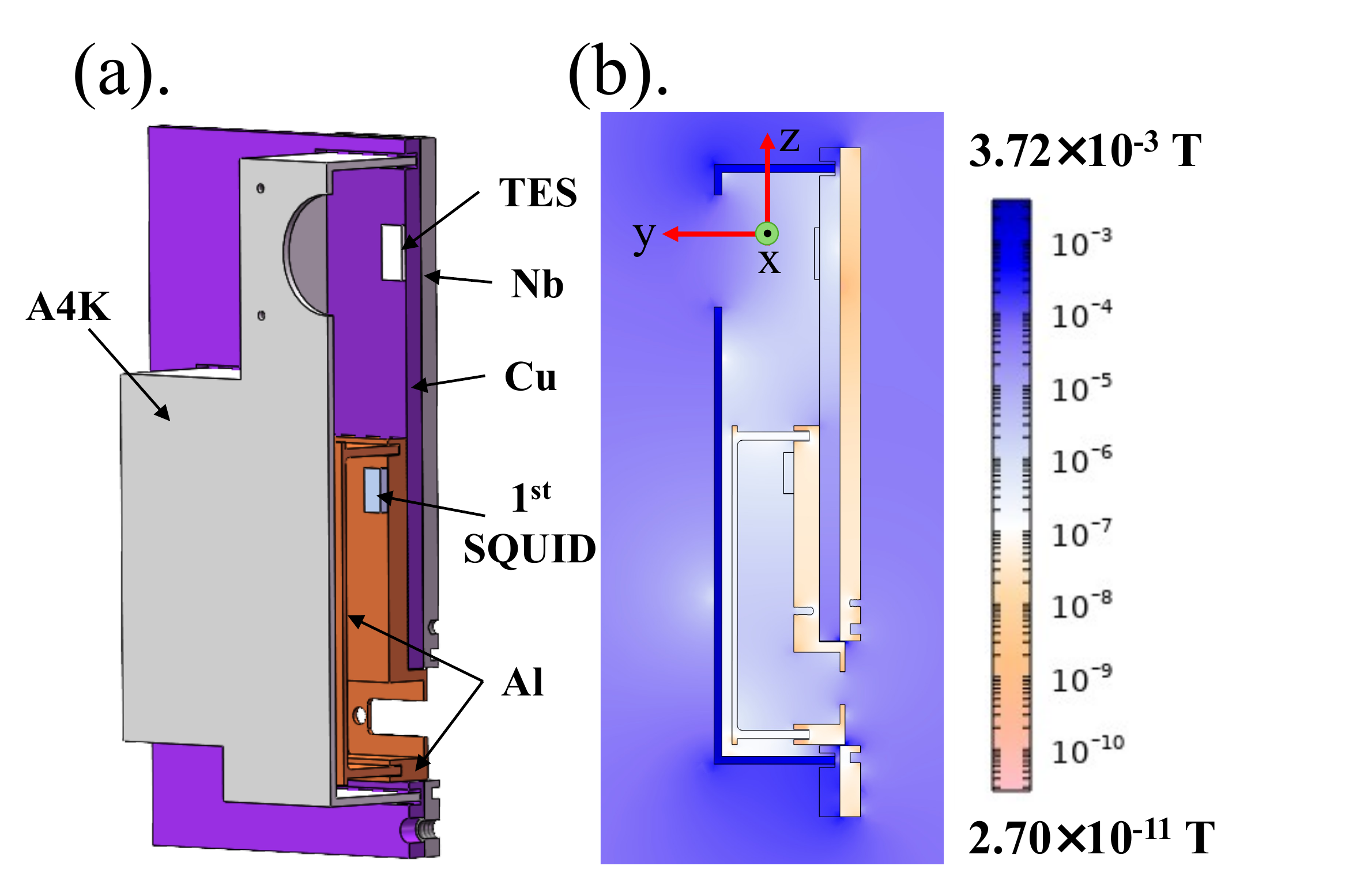} 
\caption{\label{fig12_sectional_view_of_box}
(a) Cross-sectional diagram of the test box. The test box is composed of four components: (1) The Cryoperm 10 cover plate with an incident window; (2) An almost enclosed aluminum box housing the SQUID; (3) A heat sink made of oxygen-free copper material (purple); (4) A Nb bottom plate. (b) Magnetic field distribution inside and outside the test box. The magnetic permeability of Cryoperm 10 at 100 mK was supposed to be 50000 in terms of the material's official data\cite{Cryoperm-10}. The Al and Nb as superconductor were set to have the same permeability of $10^{-5}$, and copper to 1. Based on the laboratory’s geographical location, We determined the direction and strength of the surrounding geomagnetic field: along the direction parallel to the TES (the X-axis and Z-axis), the magnetic field strength was 4 $\mu$T and 45 $\mu$T, respectively; while perpendicular to the TES (the Y-axis), the magnetic field strength was 21 $\mu$T.}
\end{figure}

\begin{figure}
\includegraphics[width=0.45\textwidth]{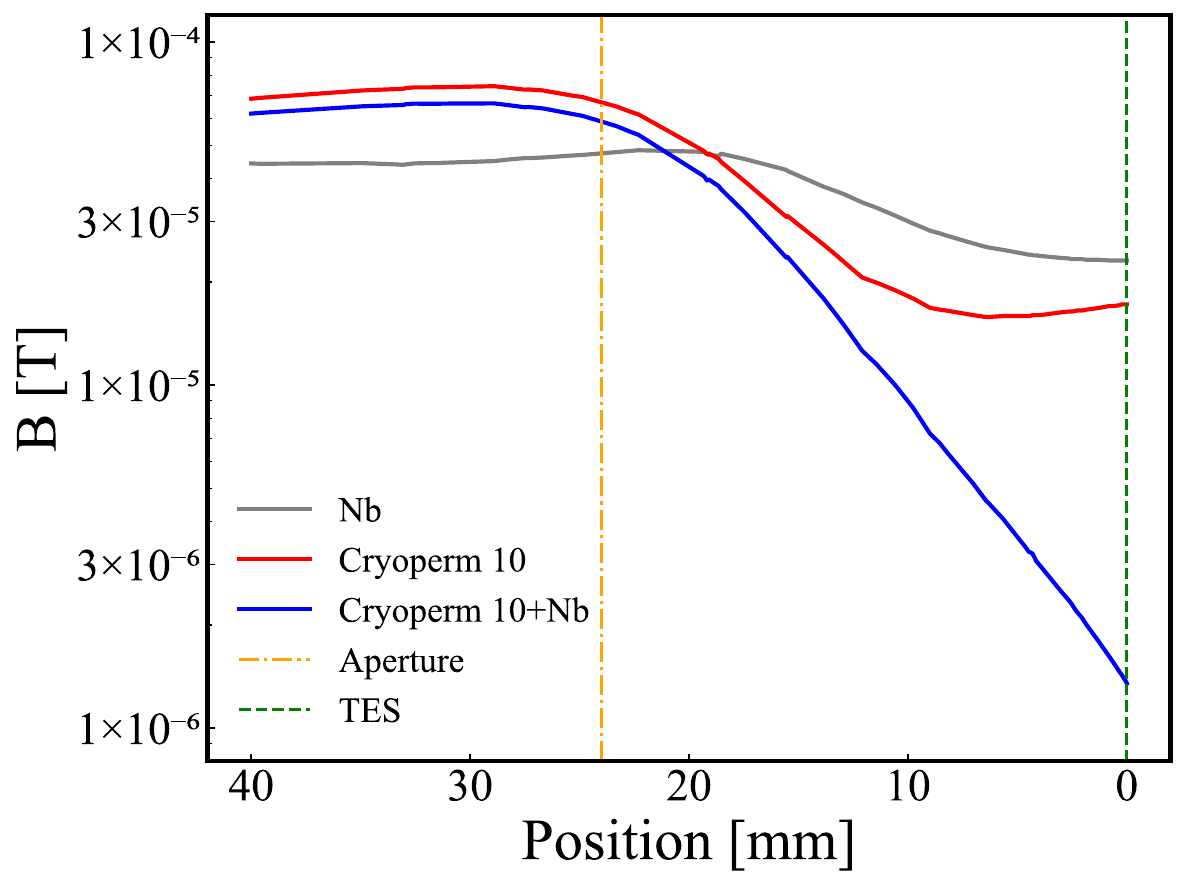} 
\caption{\label{fig13_Y-axis_magnetic_field} 
Magnetic field strength at different heights above the TES plane. The solid lines were results for three different shielding designs: (1) The top cover plate and the bottom plate were Nb metal, (2) Both were Cryoperm 10 alloy. (3) The top cover plate was Cryoperm 10 alloy and the bottom plate was Nb metal. The dashed line was the position of the TES detector and aperture.}
\end{figure}

Despite its lower magnetic susceptibility compared to bilayer TES, the AlMn TES still requires magnetic shielding to ensure stable operation, and the SQUID likewise needs shielding. We designed a dedicated magnetic shield for them, as shown in Fig.~\ref{fig12_sectional_view_of_box} (a). It mainly consisted of a 1.5 mm thick Cryoperm 10 cover plate and a 2 mm thick Nb bottom plate. The SQUID amplifier was housed inside an Al shielding enclosure and positioned parallel to the TES detector on the copper heat dissipation plate. Both were enveloped by the magnetic shield. The Cryoperm 10 cover plate features an 11-mm radius aperture directly above the TES detector to allow X-ray to pass through and be detected. The distance between the aperture and the TES detector was approximately 24 mm. In order to evaluate the effectiveness of shielding, we made several simulations based on the COMSOL Multiphysics software. Fig.~\ref{fig12_sectional_view_of_box} (b) shows that the Cryoperm 10 cover plate effectively gathers the surrounding magnetic field lines, while the superconducting Nb bottom plate obstructs their propagation. This combination effectively prevented the Earth's magnetic field from reaching the TES. Along the Y-axis through the center of the TES, we extracted the total magnetic field for three different magnetic shielding configurations, as shown in Fig.~\ref{fig13_Y-axis_magnetic_field}. The currently designed shield reduced the magnetic field near the TES detector to 1.35 $\mu$T, significantly smaller than 23.1 $\mu$T of the previous pure Nb shield and the 17.2 $\mu$T of the pure Cryoperm 10 shield. The total magnetic field at the SQUID location is 0.45 $\mu$T, achieved with the aid of an additional superconducting aluminum enclosure. The low magnetic field at the TES and SQUID locations guarantees optimal performance of both devices during operation\cite{Miniussi-2019}.

\begin{figure}
\includegraphics[width=0.45\textwidth]{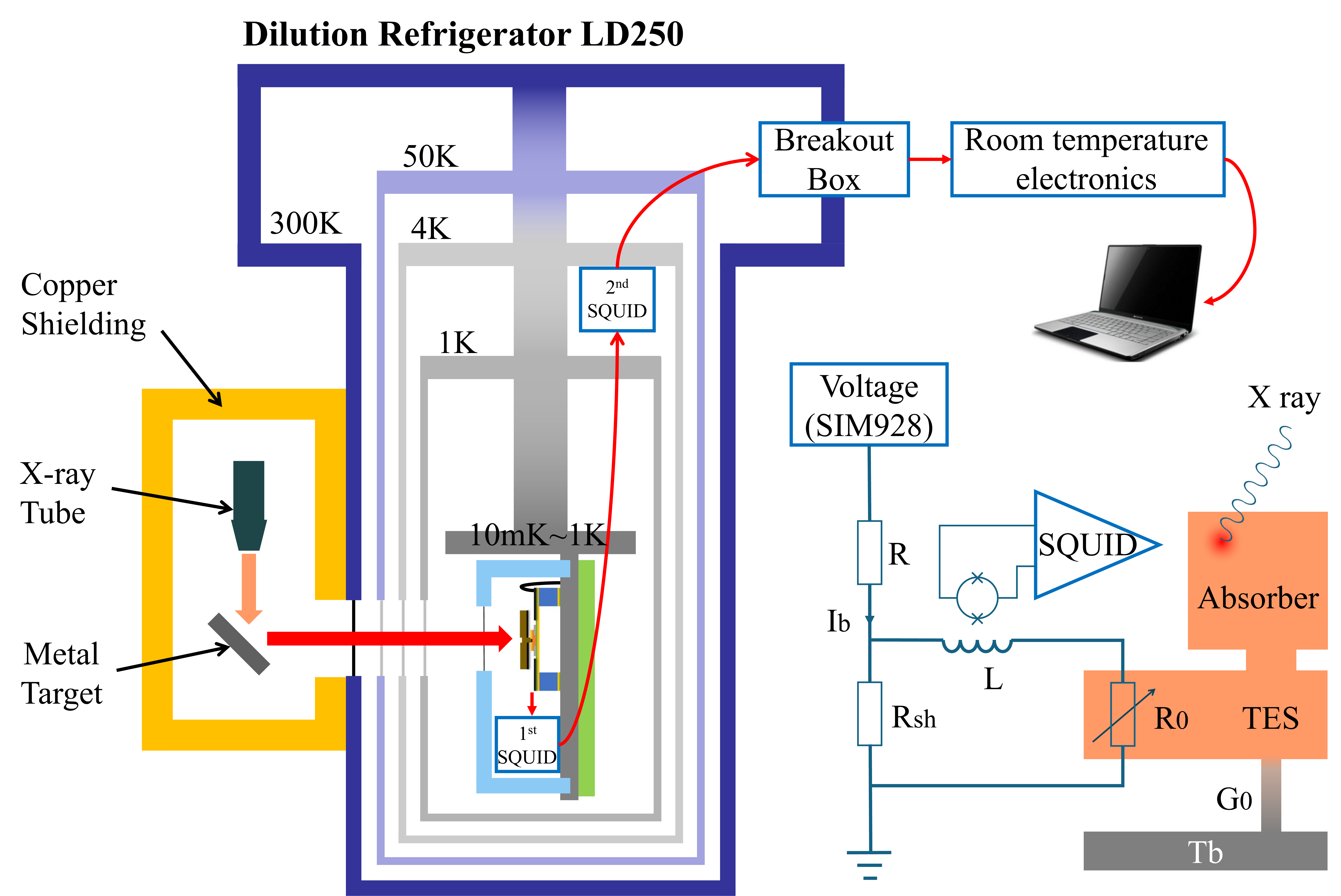} 
\caption{\label{fig14_test_sysem} 
A schematic diagram for the TES detector test system. X-ray tube was used to generate high energy photons, LD250 to provide a low temperature circumstance for the TES detector, and the right circuit to bias the TES. The window on the cryostat housing was made of 1.5 mm thick beryllium (Be), and the windows on the internal 50 K, 4 K, and 1 K shields were made of 20 $\mu$m-thick aluminum foil.} 
\end{figure}

Based on a dilution refrigerator LD250 from Bluefors and a two-stage SQUID amplifier from STAR Cryoelectronics, a testing system was constructed to characterize the TES detector, as shown in Fig.~\ref{fig14_test_sysem}. The TES detector was fixed and thermally linked to the copper heat base using Rubber Cement and numerous gold bonding wires. It was then electrically connected to the first stage of the SQUID, followed by the second stage of the SQUID located on the 4K plate, and subsequently to the room-temperature electronics with a sampling rate of 2 MHz. We used a low-noise voltage device SIM928 in series with a 1000 $\Omega$ resistor to form a current source. The shunt resistor $R_{\text{sh}}$  was 0.3 m$\Omega$. A 200 nH inductor was introduced in series with the TES to effectively suppress high-frequency noise while increasing the rise time of the TES detector signal, thereby reducing the bandwidth requirements for the SQUID. Together with the SQUID’s intrinsic input inductance of 30 nH, the total series inductance was 230 nH.

\begin{figure}
\includegraphics[width=0.45\textwidth]{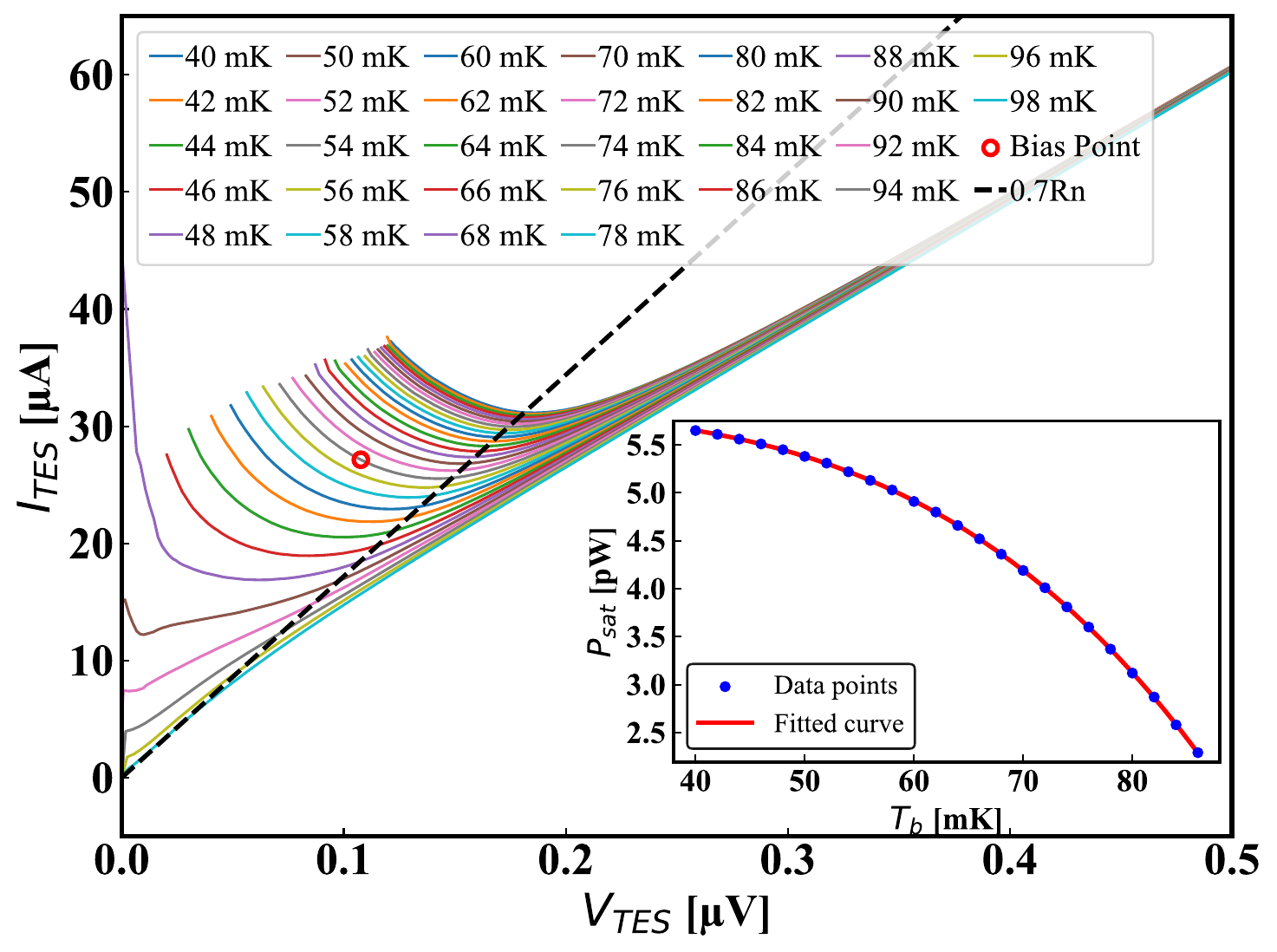} 
\caption{\label{fig15_IV_PT} 
 I-V curves of the TES detector at base temperatures from 40 mK to 98 mK. The red circle indicates the bias point, corresponding to a base temperature of 74 mK and a voltage of 0.385 V. The inset is Joule powers of the TES detector operating at 70 \% $R_n$ across various bath temperatures.}
\end{figure}

Within the temperature range of 40mK to 98 mK, we densely tested the I-V curves of the TES detector with a temperature step of 2 mK and a voltage step of 5 mV, as shown in the left panel of  Fig.~\ref{fig15_IV_PT}. Based on the I-V curves, the normal resistance of the TES was calculated to be 8.3 m$\Omega$, slightly higher than the expected value of 7.2 m$\Omega$. The current and voltage values corresponding to 70\% of Rn were extracted from the I-V curve, generating the P-T curve shown in the inset figure in Fig.~\ref{fig15_IV_PT}. The heat flow equation \( P(T) = K(T^n - T_b^n) \) was used to fit the data, yielding a superconducting critical temperature of the TES of 98.4 mK, which is close to the expected value of 100 mK. The prefactor K was $2.99 \times 10^{-8}$
 W/K$^n$, and the index n was 3.68. The thermal conductivity \( G(T) = nK\,T^{\,n-1} \) was deducted to be 220 pW/K at 98.4 mK, slightly higher than the design value of 203 pW/K predicted by the model $G=4\xi A\sigma T^3$, where A is the phonon-emitting area (taken as the product of the outer perimeter of the TES and the membrane thickess), $\xi$ is assumed to be 1, and $\sigma$=157 W/m$^2$/K$^4$ is the Stefan-Boltzmann constant given by Reference  \cite{Holmes-1998}.

\begin{figure}
\includegraphics[width=0.45\textwidth]{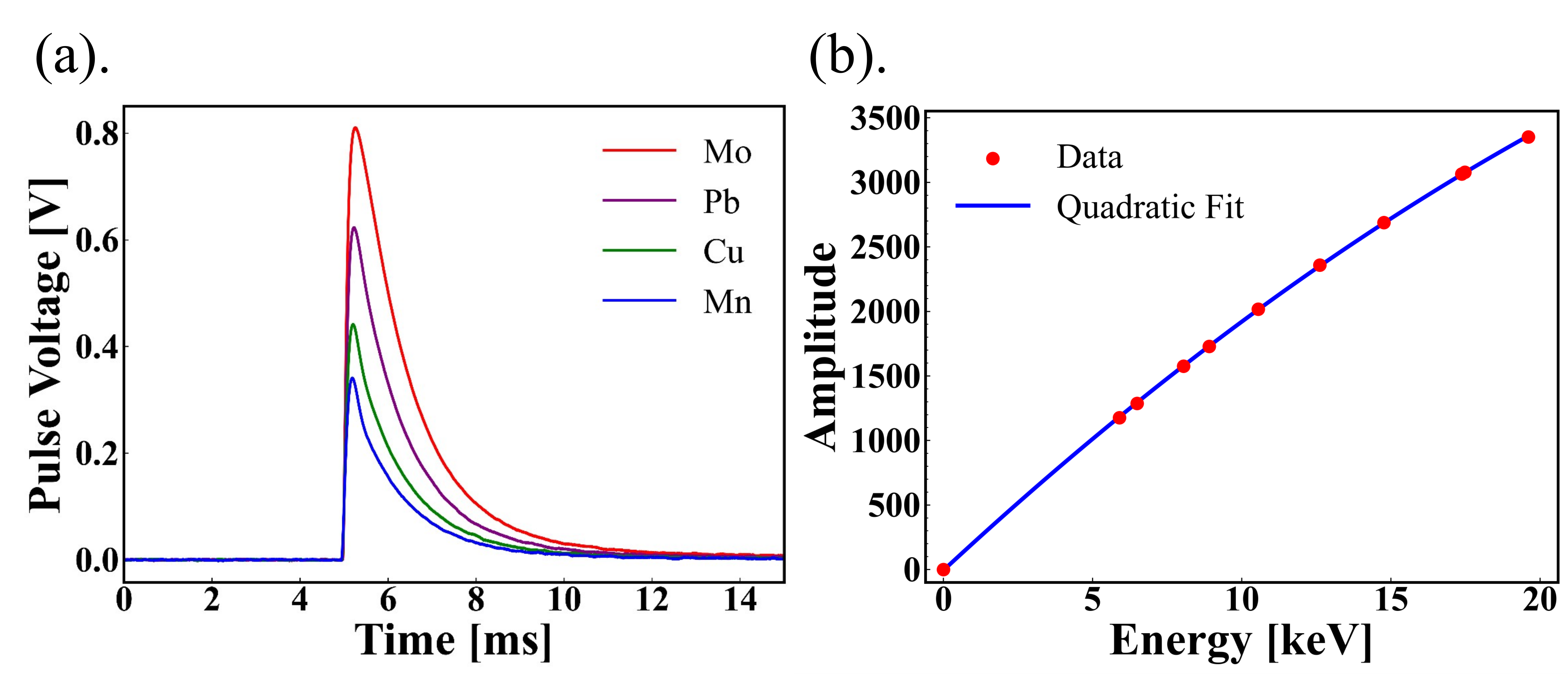}
\caption{\label{fIg16_pulse_and_quadratic_fit} 
(a) Classical pulse images of four elements.
(b) Quadratic fit of amplitude versus energy.}
\end{figure}

We selected a bias point that provides a favorable signal-to-noise ratio for testing the performance of the TES detector. The bias voltage was set to 0.385 V, and the base temperature was maintained at 74 mK. Correspondingly, the TES resistance R$_0$ was 48\% Rn. We used the bremsstrahlung radiation light generated by the Mini-X2 X-ray tube to irradiate Mn, Cu, Pb, and Mo targets respectively to produce the required characteristic X-ray photons. The X-ray tube and metal target were housed within a copper shield equipped with a collimating aperture to ensure operational safety and restrict the X-ray emission angle, thus reducing the radiative heat load on the mixing chamber of the dilution refrigerator. In addition, to prevent temperature fluctuations in the TES thermal bath due to excessive photon loading, we adjust the X-ray tube current to limit the detected count rate to approximately one event per second. Fig.~\ref{fIg16_pulse_and_quadratic_fit}(a) shows the typical pulse signals generated by the TES detector upon absorbing photons of Mn K$\alpha_1$ (5.899 keV), Cu K$\alpha_1$ (8.049 keV), Pb L$\beta$1 (12.613 keV), and Mo K$\alpha_1$ (17.480 keV). At the bias point, the temperature sensitivity $\alpha_I = \frac{T_0}{R_0} \frac{\partial R}{\partial T} $ and the current sensitivity $\beta_I = \frac{I_0}{R_0} \frac{\partial R}{\partial I} $ were calculated to be 13.7 and 0.3 respectively. Correspondingly, the loop gain $L_I = \frac{\alpha_I P_0}{G_0 I_0}$ was 2.2. The decay constant \( \tau_- = 1.15 \, \text{ms} \) was extracted by fitting the Mn K$\alpha_1$ pulse with a two-exponential function $V = V_0 (e^{-t/t_+}-e^{-t/t_-})$. According to the equation \( \tau \approx \tau_- \frac{1 + \beta_I + L_I}{1 + \beta_I} \), we obtained the intrinsic decay time \( \tau = 3.0 \, \text{ms} \), which yielded the detector's total heat capacity as 0.6 pJ/K. This value is approximately five times higher than the calculated value based on the specific heat reported in Reference\cite{Brown-2008}, and behaves similarly to the previous detector. To address this discrepancy, AlMn TES detectors without absorbers are planned for further investigation.

\begin{figure*}
\includegraphics[width=\textwidth]{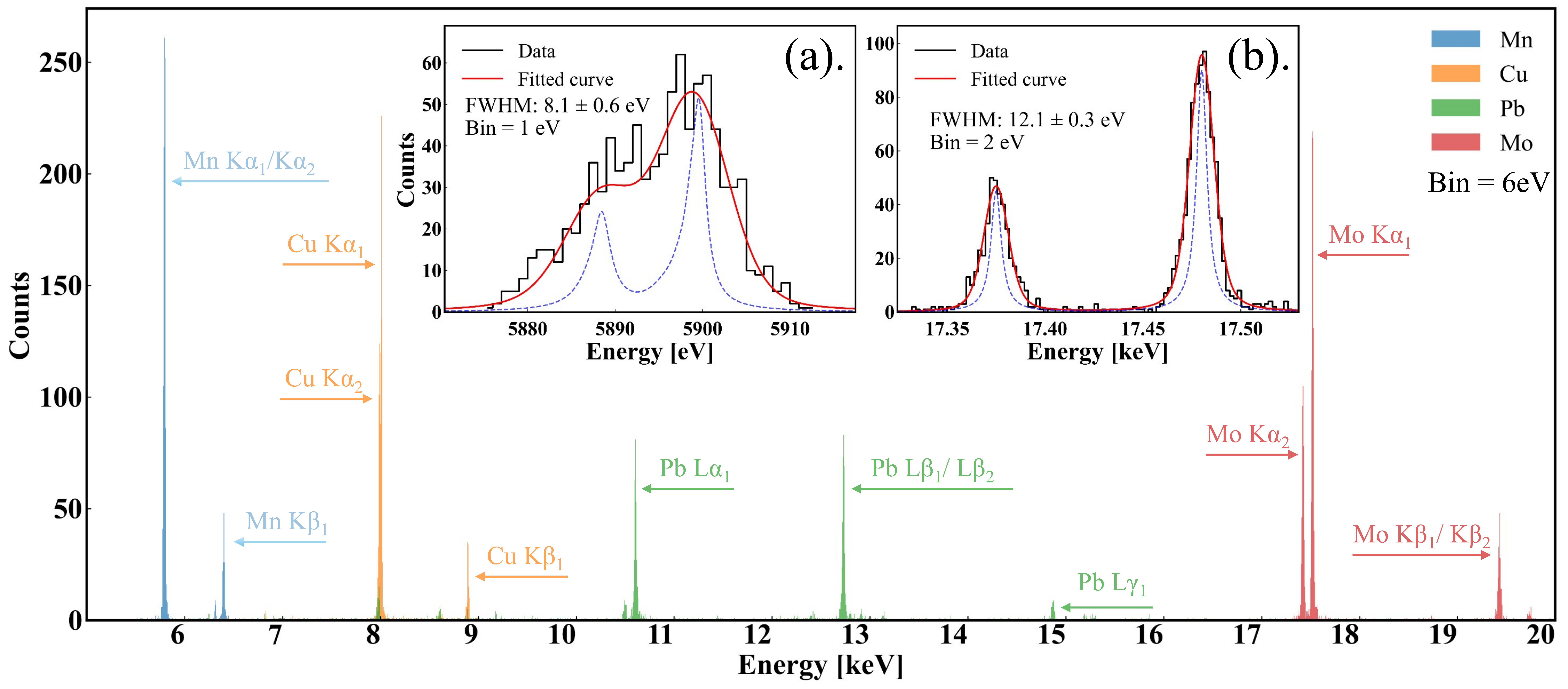}
\caption{\label{fig17_Energy_spectra}
Energy spectra for different characteristic X-ray photons generated from Mn, Cu, Pb and Mo target irradiated by high energy photons from an X-ray tube. The main characteristic X-ray lines are labeled. Inset (a) and (b) show the K$\alpha_1$ and K$\alpha_2$ lines of Mn and Mo, respectively. The dashed blue lines are natural lines shape, and the solid red curves show corresponding fits using a Voigt function.}
\end{figure*}

By applying matched filtering techniques to the raw pulse data, the filtered pulse amplitudes were extracted and plotted in Fig.~\ref{fIg16_pulse_and_quadratic_fit}(b). The relationship between pulse height and photon energy was well fitted using a quadratic function. Although the curve deviates from good linearity, it remains accurate as a calibration curve for testing unknown-energy photons. Based on this curve, we obtained a revised total energy spectrum of X-ray photons from four metal targets, as shown in Fig.~\ref{fig17_Energy_spectra}. The TES detector successfully identified the characteristic X-ray photons, such as the K$\alpha$1, K$\alpha$2 of Cu and Mo, and even the K$\alpha$1, K$\alpha$2 of Mn, as shown in the inset figures. We used a Voigt function to fit each spectrum with the natural width given by Reference\cite{Hölzer-1997,Krause1979NaturalWO} and obtained the intrinsic FWHM of the TES detector, 8.1$\pm$ 0.6 eV @ 5.9 keV, 11.4 eV$\pm$ 0.3 @ 8.0 keV, 12.1$\pm$ 0.3 eV @ 17.48 keV. The energy resolution of the TES detector was 0.69‰ at 17.48 keV. This represents the first demonstration of energy resolution beyond one thousandth in an AlMn TES X-ray detector, underscoring its potential application in X-ray spectroscopy for high-energy astronomy and material science.

\begin{figure}
\includegraphics[width=0.45\textwidth]{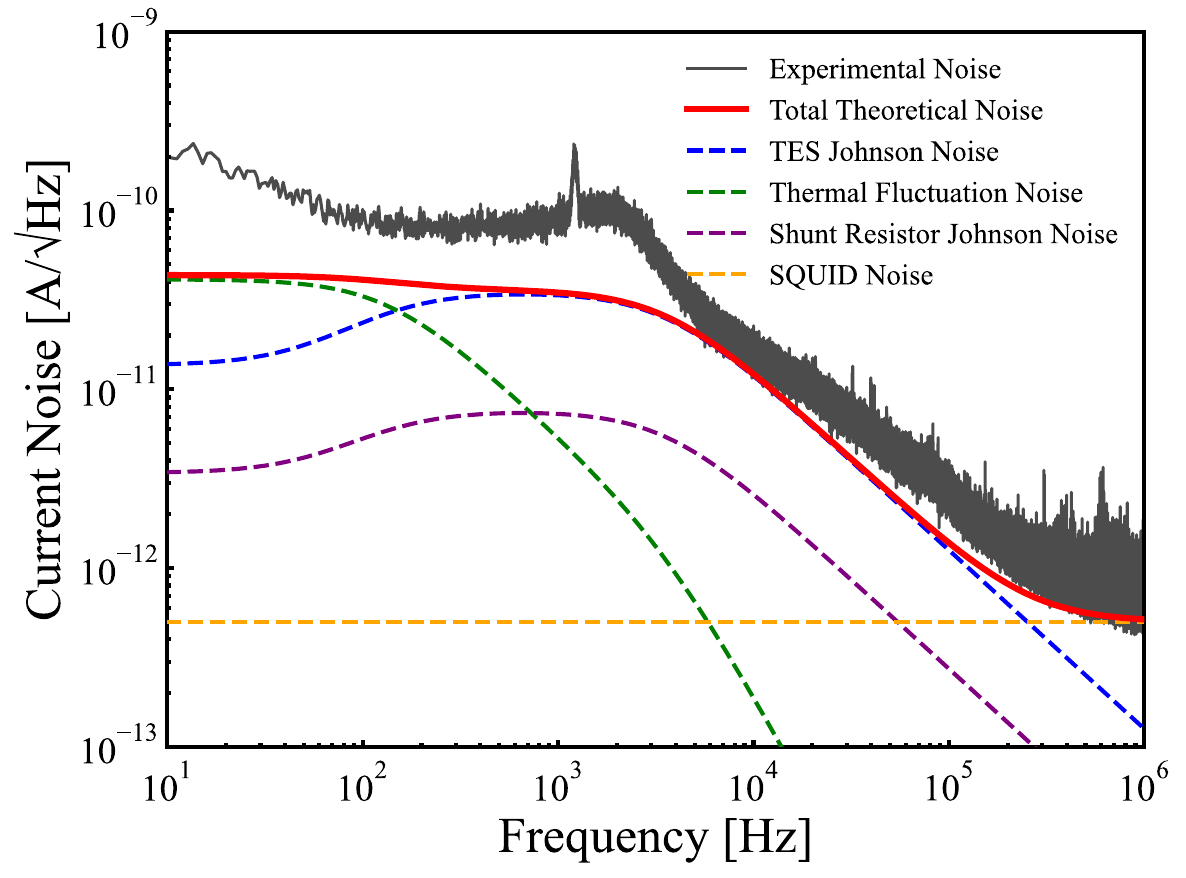} 
\caption{\label{fig18_noise} 
Measured current noise in the TES detector operated at bias point. The dashed curves show the calculated noises of four different components, in which the SQUID noise is supposed to be equal to 0.5 pA/sqrt(Hz). The black solid curve is the experimental result and the red solid curve is the total theoretical noise.}
\end{figure}
  
Fig.~\ref{fig18_noise} shows the measured total current noise of the TES detector. Using the detector parameters at the bias point and the measured current noise, the energy resolution was estimated to be 7.9 eV. This value is consistent with the measured energy resolution of 8.1 eV at 5.9 keV, but better than the 12.1 eV obtained at 17.48 keV. The degradation in resolution at higher photon energies is likely due to an increased thermal load from energetic X-rays depositing energy in the silicon substrate, which disrupts the thermal stability of the TES. 
In contrast, the fundamental limit, calculated from the sum of Johnson noise (from both the TES and the shunt resistor), thermal fluctuation noise, and SQUID noise, is approximately 3.4 eV. The discrepancy between this limit and the measured resolution indicates that excess noise dominates the observed FWHM. Future work will focus on mitigating this excess noise. Strategies include implementing electromagnetic shielding around room-temperature electronics to suppress high-frequency interference, and further reducing the X-ray beam spot size to minimize temperature fluctuations in the thermal bath. In addition, a low-pass filter will be added to the bias circuit at the mixing chamber stage to suppress noise from the SIM928 voltage source and interference coupled into the bias line. Reducing excess noise will improve the energy resolution.

In this study, we found that the tested AlMn TES exhibited a relatively small temperature sensitivity $\alpha_I$. At the operating bias point, $\alpha_I$ was 13.7; even when biased at 30\% of Rn, it only reached 23.0. One major limiting factor may be related to the annealing temperature of the AlMn film. Previous studies show that annealing above 200 °C broadens the superconducting transition \cite{Liu-2025}. Keeping the annealing temperature below 200 °C can mitigate this effect. Another contributing factor is the annular geometry of the TES. Due to its curved layout, the current density is non-uniform along the current path, which further broadens the transition compared to a rectangular TES with uniform current distribution. This geometric broadening can be reduced by increasing the inner-to-outer radius ratio, or by using a rectangle TES instead. Implementing these optimizations is expected to improve the energy resolution toward the fundamental limit of 1–2 eV, comparable to that of state-of-the-art bilayer TES detectors \cite{Smith-2012}. It is worth noting that the annular design may offer a novel strategy for tailoring the characteristics of AlMn TES devices, distinct from the conventional approach of incorporating normal-metal bars on bilayer TESs \cite{Wang-2014,Sadleir-2018}. This is a new direction that warrants further investigation.

In summary, we have designed and fabricated a novel AlMn TES detector, incorporating an Cryoperm 10 - Nb composite magnetic shield in place of the pure Nb shield. Simulation results demonstrate that this configuration reduces the geomagnetic field to 2.7\% of its ambient value, thereby preserving the performance of both the TES and the SQUID. Experimental results show that the FWHM of the pulse amplitude spectrum from the TES detector is 12.1 eV at 17.48 keV, corresponding to an energy resolution of 0.069\%. This represents the first achievement of one-thousandth energy resolution in an AlMn TES X-ray detector, indicating that this type of detector will become an important alternative in X-ray detection. This demonstrates that it is suitable for the WXPT satellite concept project. 

\begin{acknowledgments}
This work was supported by the National Natural Science Foundation of China (Grant Nos. 12275292), the National Key Research and Development Program of China (Grant Nos. 2021YFC2203402).
\end{acknowledgments}

\nocite{*}

\bibliography{sn-bibliography}

\end{document}